\documentclass[preprint,prl,12pt, superscriptaddress]{revtex4-1}
\usepackage{balance}
\usepackage[utf8]{inputenc}
\usepackage[english]{babel}
\usepackage{subcaption}
\usepackage{mathtools}
\usepackage{braket}
\usepackage{float}
\usepackage{graphicx}
\usepackage{siunitx}
\usepackage{comment}
\usepackage{url}
\usepackage{dirtytalk}
\usepackage[colorlinks = true,
            linkcolor = blue,
            urlcolor  = blue,
            citecolor = blue,
            anchorcolor = blue]{hyperref}   

\usepackage{hyperref}

\usepackage[justification=raggedright,singlelinecheck=false,labelfont=bf]{caption}  
\newcommand{\etal}{\textit{et al.~}}
\newcommand{\muunit}{cm$^2$ V$^{-1}$ s$^{-1}$}

\hypersetup{
    colorlinks=true,
}

\usepackage{scalerel}
\usepackage{tikz}
\usetikzlibrary{svg.path}
\definecolor{orcidlogocol}{HTML}{A6CE39}
\tikzset{
  orcidlogo/.pic={
    \fill[orcidlogocol] svg{M256,128c0,70.7-57.3,128-128,128C57.3,256,0,198.7,0,128C0,57.3,57.3,0,128,0C198.7,0,256,57.3,256,128z};
    \fill[white] svg{M86.3,186.2H70.9V79.1h15.4v48.4V186.2z}
                 svg{M108.9,79.1h41.6c39.6,0,57,28.3,57,53.6c0,27.5-21.5,53.6-56.8,53.6h-41.8V79.1z M124.3,172.4h24.5c34.9,0,42.9-26.5,42.9-39.7c0-21.5-13.7-39.7-43.7-39.7h-23.7V172.4z}
                 svg{M88.7,56.8c0,5.5-4.5,10.1-10.1,10.1c-5.6,0-10.1-4.6-10.1-10.1c0-5.6,4.5-10.1,10.1-10.1C84.2,46.7,88.7,51.3,88.7,56.8z};}}
\newcommand\orcidicon[1]{\href{https://orcid.org/#1}{\mbox{\scalerel*{
\begin{tikzpicture}[yscale=-1,transform shape]
\pic{orcidlogo};
\end{tikzpicture}
}{|}}}}

\usepackage{url}

\usepackage{cleveref}
\crefname{equation}{Eq.}{Eqs.}
\crefname{section}{Sec.}{Secs.}
\crefname{subsection}{Sec.}{Secs.}
\crefname{appendix}{Appendix}{Appendices}
\crefname{figure}{Figure}{Figures}
\crefname{table}{Table}{Tables}

\begin{document}

\title{Charge transport in BAs and the role of two-phonon scattering}

\author{Iretomiwa Esho \orcidicon{0000-0002-3746-6571}}
\affiliation{%
Division of Engineering and Applied Science, California Institute of Technology, Pasadena, California 91125, USA
}

\author{Austin J. Minnich \orcidicon{0000-0002-9671-9540}}
 \email{aminnich@caltech.edu}
\affiliation{%
Division of Engineering and Applied Science, California Institute of Technology, Pasadena, California 91125, USA
}%

\date{\today}

\begin{abstract}
 The semiconductor BAs has drawn significant interest due to experimental reports of simultaneous high thermal conductivity and ambipolar charge mobility. The \textit{ab~initio} prediction of high electron and hole mobility assumed the dominance of charge carrier scattering by one phonon. Recently, higher-order electron-phonon scattering processes in polar and non-polar semiconductors have been reported to have a non-negligible impact on charge transport properties, suggesting they may play a role in BAs as well. Here, we report an \textit{ab~initio} study of two-phonon electron and hole scattering processes in BAs. We find that inclusion of these higher-order processes  reduces the computed room temperature electron and hole mobility in BAs by around 40\% from the one-phonon value, resulting in an underestimate of experimental values by a similar percentage. We suggest an experimental approach to test these predictions using luminescence spectroscopy that is applicable to the defective samples which are presently available.
 
\end{abstract}
\maketitle


BAs is a semiconductor of substantial recent interest beginning from the $\it{ab~initio}$ prediction of high thermal conductivity comparable to that of diamond \cite{lindsay2013first} owing in part to the high optical phonon energy ($\sim 80$ meV) which inhibits phonon scattering. The prediction of high optical phonon energy was initially confirmed by inelastic x-ray scattering \cite{ma2016boron}, but reports of the  thermal conductivity values were significantly lower than the predictions due to scattering by As vacancies \cite{lv2015experimental}. After improvements in synthesis resulting in higher-quality samples, the high thermal conductivity was confirmed experimentally \cite{kang2018experimental, li2018high, tian2018unusual}. Further, four-phonon processes were found to make a non-negligible contribution to phonon scattering, yielding a lower thermal conductivity compared to the original predictions assuming three-phonon scattering and in quantitative agreement with experiments.

BAs has also been predicted to exhibit simultaneous high electron and hole mobilities, with computed room-temperature values of 1400 \muunit~and 2110 \muunit, respectively \cite{liu2018simultaneously, bushick2020boron}.
However, initial experiments that estimated the mobility from  conductivity and thermoelectric measurements and a single parabolic band model yielded a lower hole mobility of 400 \muunit~\cite{kim2016thermal}; recent direct Hall measurements  yielded $\sim$ 500 \muunit~on bulk samples \cite{shin2022high}. The lower values obtained experimentally have been attributed to scattering by charged impurities in the defective samples which could be synthesized. Recent experiments have circumvented the need for high-quality macroscopic samples by measuring the ambipolar diffusivity of photoexcited carriers in a local region of the sample using transient grating experiments \cite{shin2022high} or transient reflectivity microscopy \cite{yue2022high}. Using the Einstein relation to convert the measured diffusivity into a mobility, these experiments obtained ambipolar mobilities of 1500--1600 \muunit~at some locations on the sample. These values are in good agreement with those calculated from first principles \cite{liu2018simultaneously}.

Most first-principles studies of the electron-phonon interactions employ the lowest level of perturbation theory involving one electron and one phonon (1ph) \cite{bernardi2014ab, giustino2017electron}, and this level of theory was also used for BAs \cite{liu2018simultaneously, bushick2020boron}. Given the contribution of higher-order phonon processes to thermal transport in BAs, it is of interest to consider the role of higher-order processes in charge transport. Although evidence for the contribution of multiphonon processes to electron-phonon scattering has been previously reported \cite{sher1967resonant, ngai1972two, alldredge1967role}, only recently have first-principles studies included the contribution of higher-order scattering processes, such as that of an electron with two phonons (2ph) in the electron-phonon interaction \cite{lee2020ab, cheng2022high, hatanpaa2023two}. In GaAs at room temperature, the 2ph scattering rates were predicted to be on the order of the 1ph rates \cite{lee2020ab}, resulting in a $\sim$ 40\% reduction to the computed mobility at 300 K. Good quantitative agreement with experimental mobility was obtained only considering this correction. Corrections to the high-field transport properties of GaAs of a similar magnitude were also found \cite{cheng2022high}. For non-polar semiconductors, Hatanp{\"a}{\"a} \etal reported improved agreement of the warm electron coefficient in Si  over temperatures from 190 K to 310 K with the inclusion of 2ph processes \cite{hatanpaa2023two}. These studies suggest that inclusion of 2ph processes for electron-phonon scattering may be necessary to accurately predict the charge transport properties of semiconductors.  

Here, we report an \textit{ab~initio} study of the role of two-phonon scattering of electrons and holes in  BAs. We find that the two-phonon rates may be as large as $\sim 50$\% of the one-phonon rates, leading to a marked reduction in the  calculated ambipolar mobility from 1420 \muunit~to 810 \muunit~at room temperature and a 35\%--50\% correction to the carrier mobility over temperatures from 150 K to 350 K. The experimental origin of the discrepancy could arise from the super-diffusion of hot carriers shortly after photoexcitation, an effect which has been observed using scanning ultrafast electron microscopy, leading  to an overestimate of the ambipolar diffusivity. On the  theory side, an underestimate of the predicted value is possible owing to cancellation between the iterated and direct contributions to 2ph scattering, the latter of which is neglected here. To test our predictions given the defective samples presently available, we suggest an experimental approach based on direct measurements of hot carrier lifetimes using the broadening of photoluminescence spectra.


We computed the mobility of electrons and holes in BAs using established methods based on DFT and DFPT \cite{choi2021electronic, ponce2021first, bernardi2016first, giustino2017electron}. Briefly, we obtained the electronic structure and electron-phonon matrix elements using \textsc{Quantum Espresso} \cite{giannozzi2009quantum} with a relaxed lattice constant of $\SI{4.819}{\angstrom}$, a coarse $12 \times 12 \times 12$ \textbf{k} grid, and plane wave cutoff of 80 Ry. A fully relativistic ultrasoft potential with the Perdew–Burke–Ernzerhof (PBE) exchange-correlation functional was used. For the DFPT calculations, we employed a $6 \times 6 \times 6$ phonon \textbf{q} grid. The band structure and electron-phonon matrix elements were interpolated onto a fine 160$^3$ and 80$^3$ \textbf{k} and \textbf{q} grid, respectively,  using \textsc{Perturbo} \cite{zhou2021perturbo}. Increasing the grid density to 200$^3$ and 100$^3$ for the \textbf{k} and \textbf{q} grids, respectively, changed the mobility by 2\%. The Fermi level was chosen so as to obtain a carrier concentration of 10$^{15}$ cm$^{-3}$ at all temperatures. The energy window was set to 200 meV above (below) the band extremum for electrons (holes). Increasing the energy window to 250 meV changed the mobility by 0.6\%. We explicitly constructed the collision matrix and solved the Boltzmann transport equation using numerical linear algebra, from which transport properties were calculated. Details of this approach are given elsewhere \cite{choi2021electronic, cheng2022high, hatanpaa2023two}. The contributions of the next-to-leading order electron-phonon scattering processes originally derived in \cite{lee2020ab} were computed following the implementation used in \cite{cheng2022high}. The 2ph rates were iterated five times. Increasing the number of iterations to six changed the mobility by 2.7\%.

\begin{figure}
    \centering
    {
    \includegraphics[width=0.9\textwidth]{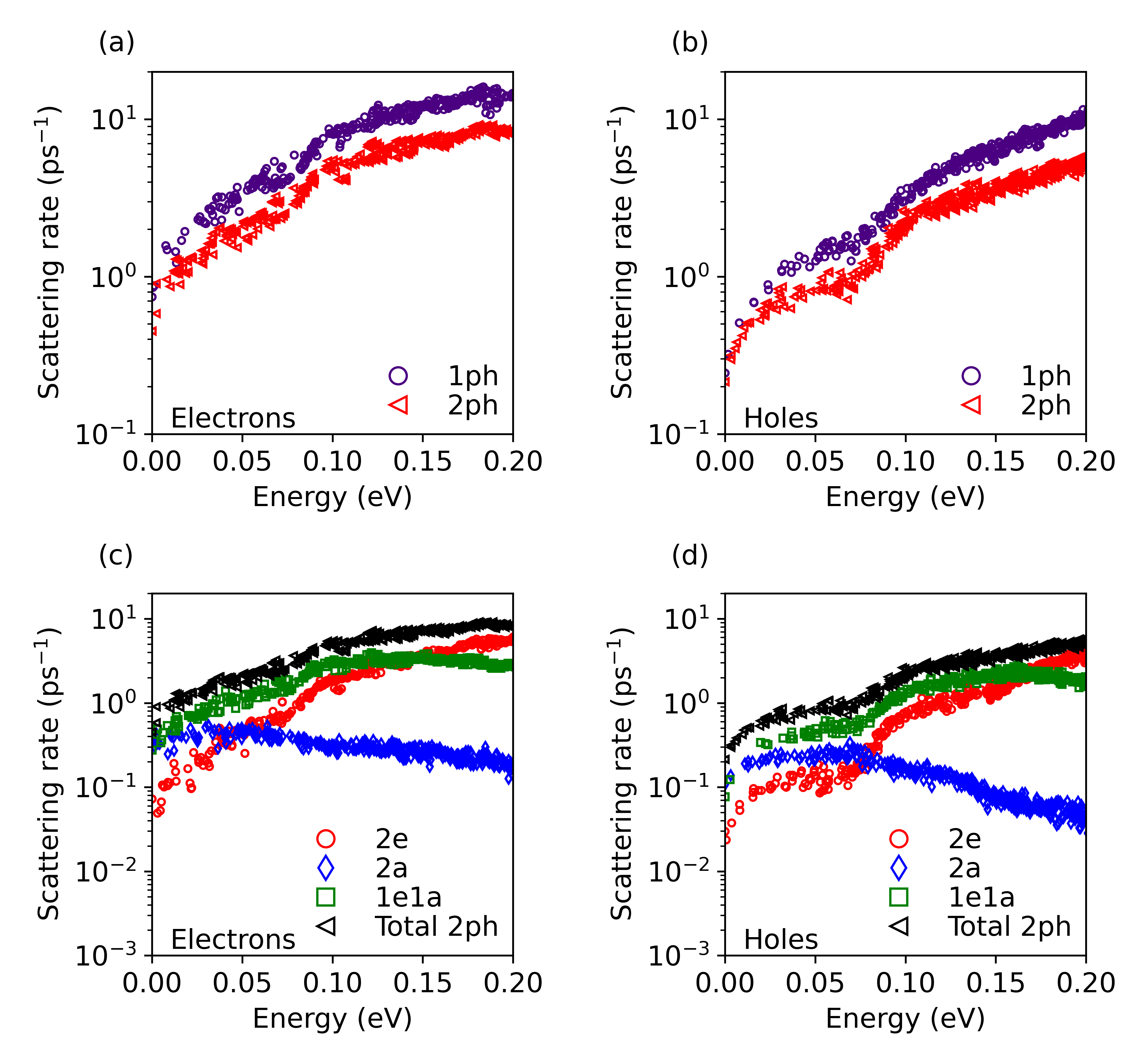}
    \phantomsubcaption\label{fig:rate_e} \phantomsubcaption\label{fig:rate_h}
    \phantomsubcaption\label{fig:2phrate_e} \phantomsubcaption\label{fig:2phrate_h}
    }
    \caption{Scattering rates versus energy for (a) electrons and (b) holes in BAs including  1ph (circles) and 2ph processes (triangles) at 300 K. The computed 2ph rates for electrons and holes are around 50\% of the 1ph rates. Computed total 2ph (triangles), 1e1a (squares), 2a (diamonds), and 2e (circles) scattering rates versus energy for (c) electrons and (d) holes show the sub-processes that comprise the total 2ph rates. Below 150 meV, the 1e1a processes have the largest contribution to the 2ph rates at 300 K. }
    \label{fig:rates}
\end{figure}

The calculated scattering rates for electrons and holes are shown in \cref{fig:rate_e,fig:rate_h}. The trend of the 1ph scattering rates agrees with that reported previously \cite{liu2018simultaneously}; quantitative differences are due to differing exchange-correlation functional or pseudopotential  necessitated by the use of \textsc{Perturbo} in this work. We observe the characteristic sharp increase in the scattering rate for electrons and holes near $\hbar \omega _\text{LO} \sim 80$ meV as LO-phonon emission starts to dominate the electron-phonon interaction. The 2ph rates largely follow the same trend and are on the order of  the 1ph rates, consistent with previously published 2ph calculations for GaAs \cite{cheng2022high, lee2020ab} and Si \cite{hatanpaa2023two}. At 300 K, the 2ph rates are around 50\% of the 1ph rates. Prior works have examined the influence of the exchange-correlation functional on charge carrier mobilities, finding variations on the order of $\sim 10-15$\% in Si \cite{ponce2018towards} and BAs \cite{liu2018simultaneously}. Although this uncertainty may influence the predicted absolute mobility values, we expect the relative contribution of 2ph processes compared to 1ph processes to be insensitive to the choice of functional.

2ph processes exhibit several different sub-types because the two phonons involved in scattering can each be  emitted or absorbed. Following Ref.~\cite{lee2020ab}, processes where a phonon is emitted and another absorbed are denoted 1e1a, and processes where two phonons are sequentially emitted or absorbed are 2e and 2a, respectively. The individual sub-processes contributing to the total 2ph rate are shown in \cref{fig:2phrate_e,fig:2phrate_h} for electrons and holes, respectively. Below $\hbar \omega _\text{LO} \sim $ 80 meV, 1e1a processes are dominant. Note that the total 1e1a rate includes processes where a phonon is first emitted and another absorbed, and processes where a phonon is first absorbed and another  is subsequently emitted. Two-phonon emission (2e) processes are comparatively weak in this region since LO phonon emission is prohibited until the energy threshold of $2 \hbar \omega_\text{LO}$.
Two-phonon absorption (2a) processes are generally weak  throughout the energy range studied, except at sufficiently low energies where emission and therefore 1e1a events become increasingly unlikely such that 2a rates are comparable to 1e1a rates. Between $\hbar \omega _\text{LO}$ and 2$\hbar \omega _\text{LO}$, the 1e1a and 2e rates increase as LO phonon emission starts to dominate the electron-phonon interaction. Beyond 2$\hbar \omega _\text{LO}$, carriers are energetic enough to emit two LO phonons, and 2e processes have the largest contribution to the total 2ph scattering rate. This energy dependence of the individual 2ph sub-processes in BAs is consistent with those reported for GaAs and Si \cite{lee2020ab, cheng2022high, hatanpaa2023two}.

\begin{figure}

    \centering
    {
    \includegraphics[width=0.95\textwidth]{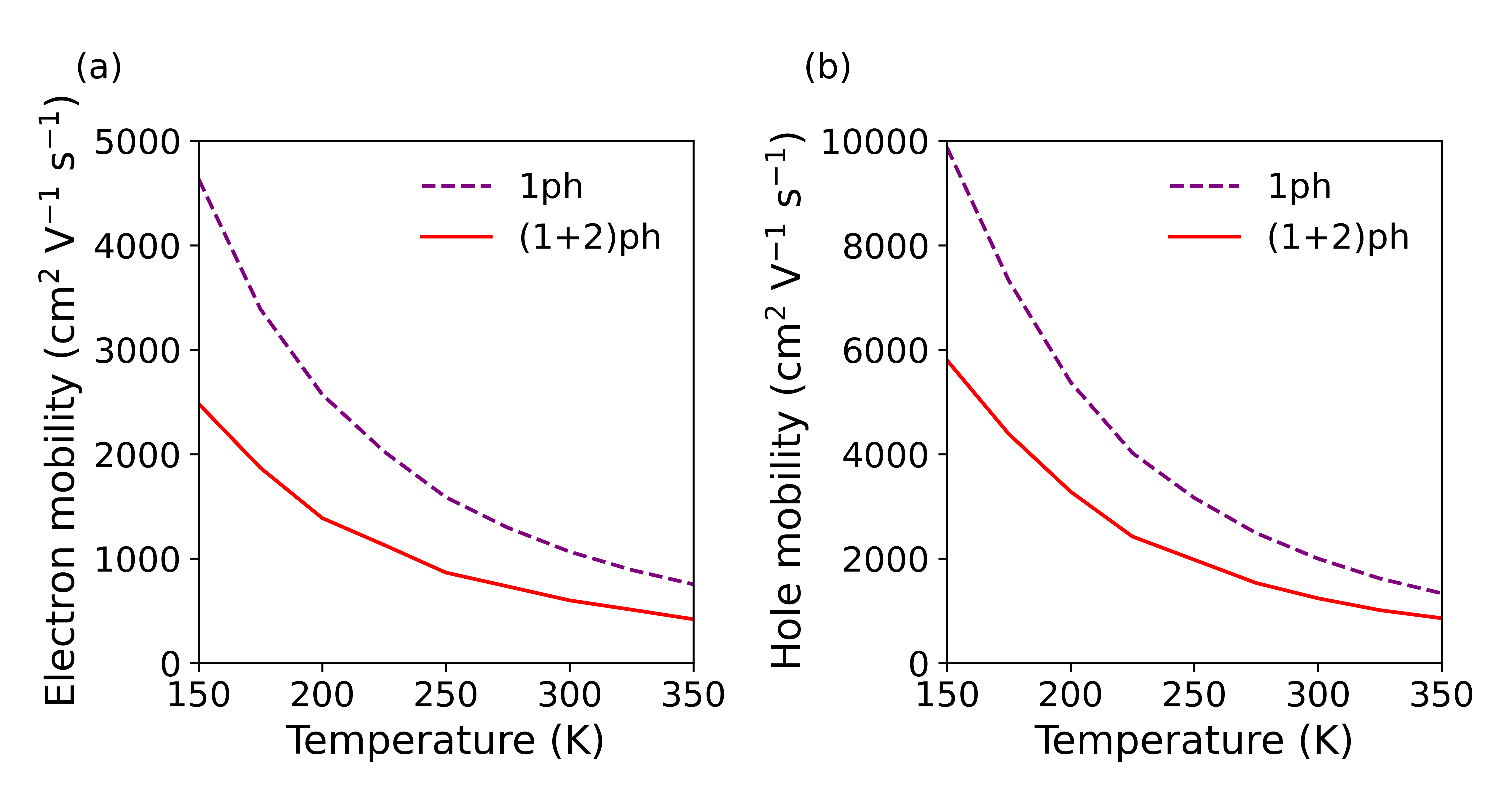}
    \phantomsubcaption\label{fig:mu_e} \phantomsubcaption\label{fig:mu_h}
    }
    \caption{(a) Electron and (b) hole mobility in BAs versus temperature at the 1ph (dashed line) and (1+2)ph (solid line) level of theory. For holes, the correction to the mobility at room temperature from including 2ph processes is $\sim$ 37\%, while for electrons this correction is $\sim$ 43\%, demonstrating the significant contribution of 2ph processes to the mobility at room temperature. }
   \label{fig:mob}

\end{figure}

We next examine the effect of 2ph processes on the electron and hole mobility. The computed 1ph and (1+2)ph mobility versus temperature is shown in  \cref{fig:mu_e,fig:mu_h} for electrons and holes, respectively. With only 1ph processes, we obtain room-temperature electron mobility $\mu_e = 1066$ \muunit~and hole mobility $\mu_h = 2000$ \muunit, in quantitative agreement with previous 1ph predictions that employ the same PBE exchange-correlation functional (see Supplementary Information of Ref.~\cite{liu2018simultaneously} for calculations using the same functional as in this work). With the inclusion of 2ph processes, $\mu_e$ and $\mu_h$ decrease to 600 \muunit~and 1240 \muunit, respectively, corresponding to a 43\% and 37\% reduction at room temperature. Over the temperature range from 150 -- 350 K, this correction ranges from 36\% at 350 K to 41\% at 150 K for holes, and 44\% at 350 K to 46\% at 150 K for electrons. These corrections to the electron mobility are of a comparable magnitude to those obtained for GaAs ($\sim$ 45\%) \cite{cheng2022high, lee2020ab}, but slightly higher than those for Si ($\sim$ 35\%) \cite{hatanpaa2023two}. 

BAs exhibits several distinct features compared to other polar semiconductors such as GaAs. In GaAs and other polar materials, LO phonons make the overwhelming contribution to electron-phonon scattering \cite{zhou2016ab}. In BAs, carrier scattering relevant to mobility is instead primarily due to acoustic phonons owing to the high optical phonon energy (80 meV versus 35 meV in GaAs) that limits scattering by LO phonon emission as well as the decreased LO phonon absorption scattering from decreased thermal population \cite{liu2018simultaneously}. Additionally, in GaAs, intervalley processes have a negligible effect on low-field charge transport because of the $\Gamma-L$ energy separation of 300 meV, but scattering processes in BAs are more similar to those in Si in that they involve intervalley transfers mediated by zone-edge wave vector phonons. 
Our calculations reveal that intervalley processes  account for 43\% of (1+2)ph scattering in BAs at 300 K and 20\% at 150 K. The decrease with decreasing temperature occurs due to reduced population of zone-edge phonons required for intervalley scattering. As a comparison, intervalley processes account for 61\% of (1+2)ph scattering in Si at 300 K and 25\% at 150 K.

We  consider our calculated mobility values in context of recent optical experiments on BAs that reported an ambipolar carrier mobility \cite{shin2022high, yue2022high}. At the 1ph level of theory, we predict a high ambipolar mobility $\mu_a = 2\mu_e \mu_h / (\mu_e + \mu_h)$ of 1420 \muunit~at 300 K using 1ph theory, consistent with a prior computed value of 1570 \muunit~with the PBE exchange-correlation functional \cite{liu2018simultaneously} and in agreement with recent experimental reports \cite{yue2022high, shin2022high}. Including 2ph processes reduces $\mu_a$ to 810 \muunit, a 43\% reduction. Considering the (1+2)ph mobility value, the apparent agreement between theory and experiment is substantially degraded, with the experiment now overestimating the theory.

This discrepancy could arise from several factors. First, the quantity that was measured in the optical experiments of Refs.~\cite{yue2022high, shin2022high} was the ambipolar diffusion coefficient of photoexcited charge carriers, from which the mobility was  obtained through the Einstein relation.  In Refs.~\cite{shin2022high, yue2022high}, the photoexcitation wavelength for determination of the ambipolar diffusion coefficient was chosen to be around the available estimates of the bandgap energy ($\sim$ 2 eV \cite{kang2019basic, buckeridge2019electronic, lyons2018impurity, bushick2019band}).  If the photon energy exceeds the bandgap energy, the photoexcited carriers will have energy in excess of thermal energies, potentially causing the extracted transport properties to differ from their linear-response values.  This hot-carrier effect was observed in both Refs.~\cite{shin2022high, yue2022high} as a larger measured electronic diffusivity for pump wavelengths $\lesssim 500$ nm.  Evidence for the absence of the hot carrier effect for the final reported diffusivity values was presented, for example, in Fig.~1D of Ref.~\cite{shin2022high}, as the plateau of the measured electronic decay rate with increasing wavelength. On the other hand, scanning ultrafast electron microscopy (SUEM) studies have reported observations of super-diffusion of photoexcited carriers in semiconductors persisting over hundreds of picoseconds \cite{najafi2017super, ruzicka2010hot, choudhry2022persistent}. This phenomenon has been attributed to the additional contribution to carrier diffusion of a pressure gradient in the non-degenerate hot carrier gas after photoexcitation \cite{najafi2017super}. In Refs.~\cite{shin2022high, yue2022high}, the diffusivity was extracted from the electronic decay curve over timescales from tens to hundreds of picoseconds, conceivably leading to an extracted diffusivity that was influenced by super-diffusion.


On the theory side, a possible cause of an underestimate for the computed mobility is the cancellation of the two contributions to electron-phonon scattering at second order. These 2ph processes may arise from the 1ph term, corresponding to the first derivative of the interatomic potential with respect to lattice displacements taken to second order in perturbation theory, or a direct 2ph term involving the simultaneous interaction of an electron with two phonons with a strength given by second-order derivative of the interatomic potential \cite{holstein1959theory, kocevar1980multiphonon}. In this work and other recent $\it{ab~initio}$ studies of 2ph scattering, only the first term was included. However, in the long-wavelength acoustic phonon limit, these two terms cancel owing to translational invariance of the crystal, and thus neglect of the second term will lead to an overestimate of 2ph scattering rate. This cancellation has long complicated the study of 2ph scattering in semiconductors \cite{ngai1974carrier, kocevar1980multiphonon}. A recent study of 2ph scattering in Si suggested that the correction could be on the order of 10-20\% in that material \cite{hatanpaa2023two}. It is possible that this effect could lead to an  underestimate of the computed mobility in BAs; further study is needed to investigate this hypothesis.

\begin{figure}[H]
     \centering
         \includegraphics[width=0.55\textwidth]{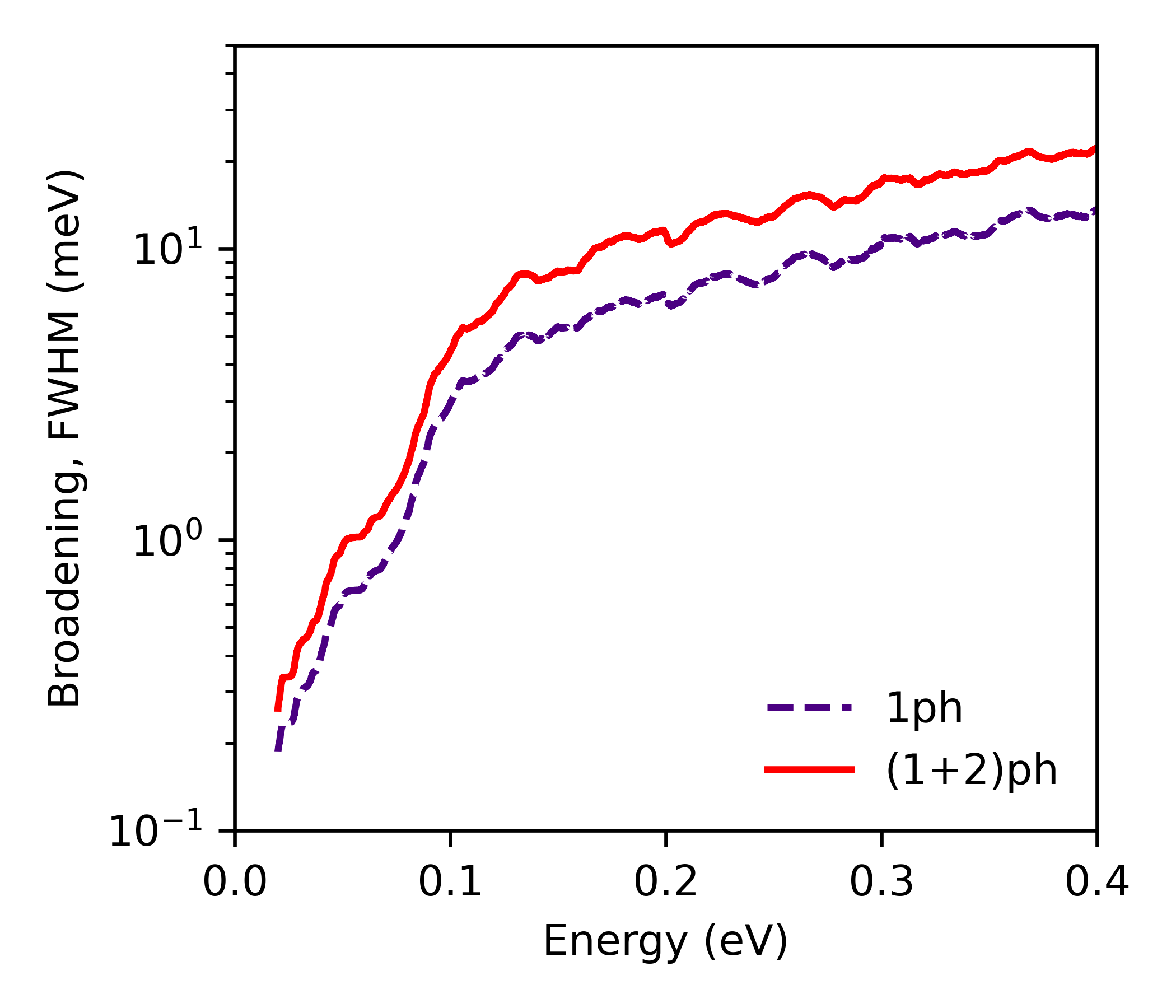}
         \caption{Calculated  broadening versus energy for  electrons due to electron-phonon scattering at 77 K and carrier concentration of 10$^{15}$ cm$^{-3}$. The difference in broadening between the 1ph (dashed line) and (1+2)ph (solid line) theory is expected to be distinguishable considering prior reports of experimental uncertainties $\sim 1$ meV \cite{fasol1990continuous}.  }
         \label{fig:broad}
         
\end{figure}

Absent higher-quality samples, verifying the prediction of the role of 2ph scattering using transport measurements is challenging due to the contribution of extrinsic defect scattering. We suggest an alternative approach based on continuous wave luminescence spectroscopy which allows the lifetimes of electronic states away from the band minimum to be determined \cite{fasol1990continuous}. These higher-energy states are less influenced by impurity scattering compared to those near the band edge, and so insight into the role of 2ph scattering can be obtained by comparing  measured photoluminescence linewidths to theory. In these experiments, hot electrons excited by a continuous-wave laser emit photons by recombination, and the spectrum of the emitted light exhibits a broadening that is determined by the lifetime of the state.
We may predict the difference in broadening at the 1ph and (1+2)ph levels of theory in BAs using the same $\it{ab~initio}$ theory employed for transport calculations. In Fig.~\ref{fig:broad}, we plot the predicted full-width at half-maximum (FWHM) of the luminescence peak, $2 \Gamma = \tau^{-1}$, versus energy for electrons. At 0.4 eV above the conduction band minimum (CBM), we predict $2 \Gamma \sim 13$ meV and 21 meV for 1ph and  (1+2)ph, respectively. This 8 meV difference is almost an order of magnitude higher than the experimental uncertainty reported in Ref.~\cite{fasol1990continuous} and thus should be discernible.

In summary, we have reported $\it{ab~initio}$ calculations of ambipolar mobility in BAs considering 2ph electron-phonon processes. We find that the inclusion of these processes reduces the predicted electron and hole mobility by 43\% and 37\% at room temperature, respectively, lowering the ambipolar mobility by 43\% and underestimating experimental reports by a similar amount. We hypothesize that the discrepancy between our results and recent optical experiments could in part arise from the super-diffusion of hot carriers, or an underestimation of the calculated mobility owing to cancellations at second-order of perturbation theory. We have suggested an experimental approach based on hot-electron luminescence to test these predictions.

I.E. was supported by the National Science Foundation Graduate Research Fellowship under Grant No. DGE-1745301. A.J.M was supported by AFOSR under Grant No. FA9550-19-1-0321. The authors thank Benjamin Hatanp{\"a}{\"a} for useful discussions and providing data on intervalley scattering in Si.

\bibliographystyle{is-unsrt}
\bibliography{refs}

\begin{thebibliography}{10}
\ifx \showCODEN  \undefined \def \showCODEN #1{CODEN #1}  \fi
\ifx \showISBN   \undefined \def \showISBN  #1{ISBN #1}   \fi
\ifx \showISSN   \undefined \def \showISSN  #1{ISSN #1}   \fi
\ifx \showLCCN   \undefined \def \showLCCN  #1{LCCN #1}   \fi
\ifx \showPRICE  \undefined \def \showPRICE #1{#1}        \fi
\ifx \showURL    \undefined \def \showURL {URL }          \fi
\ifx \path       \undefined \input path.sty               \fi
\ifx \ifshowURL \undefined
     \newif \ifshowURL
     \showURLtrue
\fi

\bibitem{lindsay2013first}
Lucas Lindsay, DA~Broido, and TL~Reinecke.
\newblock First-principles determination of ultrahigh thermal conductivity of
  boron arsenide: A competitor for diamond?
\newblock {\em Physical review letters}, 111\penalty0 (2):\penalty0 025901,
  2013.
\newblock
  {\url{https://journals.aps.org/prl/abstract/10.1103/PhysRevLett.111.025901}}.

\bibitem{ma2016boron}
Hao Ma, Chen Li, Shixiong Tang, Jiaqiang Yan, Ahmet Alatas, Lucas Lindsay,
  Brian~C Sales, and Zhiting Tian.
\newblock Boron arsenide phonon dispersion from inelastic x-ray scattering:
  Potential for ultrahigh thermal conductivity.
\newblock {\em Physical Review B}, 94\penalty0 (22):\penalty0 220303, 2016.
\newblock
  {\url{https://journals.aps.org/prb/abstract/10.1103/PhysRevB.94.220303}}.

\bibitem{lv2015experimental}
Bing Lv, Yucheng Lan, Xiqu Wang, Qian Zhang, Yongjie Hu, Allan~J Jacobson,
  David Broido, Gang Chen, Zhifeng Ren, and Ching-Wu Chu.
\newblock Experimental study of the proposed super-thermal-conductor: Bas.
\newblock {\em Applied Physics Letters}, 106\penalty0 (7):\penalty0 074105,
  2015.
\newblock {\url{https://aip.scitation.org/doi/10.1063/1.4913441}}.

\bibitem{kang2018experimental}
Joon~Sang Kang, Man Li, Huan Wu, Huuduy Nguyen, and Yongjie Hu.
\newblock Experimental observation of high thermal conductivity in boron
  arsenide.
\newblock {\em Science}, 361\penalty0 (6402):\penalty0 575--578, 2018.
\newblock {\url{https://www.science.org/doi/full/10.1126/science.aat5522}}.

\bibitem{li2018high}
Sheng Li, Qiye Zheng, Yinchuan Lv, Xiaoyuan Liu, Xiqu Wang, Pinshane~Y Huang,
  David~G Cahill, and Bing Lv.
\newblock High thermal conductivity in cubic boron arsenide crystals.
\newblock {\em Science}, 361\penalty0 (6402):\penalty0 579--581, 2018.
\newblock {\url{https://www.science.org/doi/full/10.1126/science.aat8982}}.

\bibitem{tian2018unusual}
Fei Tian, Bai Song, Xi~Chen, Navaneetha~K. Ravichandran, Yinchuan Lv, Ke~Chen,
  Sean Sullivan, Jaehyun Kim, Yuanyuan Zhou, Te-Huan Liu, Miguel Goni, Zhiwei
  Ding, Jingying Sun, Geethal Amila Gamage~Udalamatta Gamage, Haoran Sun,
  Hamidreza Ziyaee, Shuyuan Huyan, Liangzi Deng, Jianshi Zhou, Aaron~J.
  Schmidt, Shuo Chen, Ching-Wu Chu, Pinshane~Y. Huang, David Broido, Li~Shi,
  Gang Chen, and Zhifeng Ren.
\newblock Unusual high thermal conductivity in boron arsenide bulk crystals.
\newblock {\em Science}, 361\penalty0 (6402):\penalty0 582--585, 2018.
\newblock {\url{https://www.science.org/doi/full/10.1126/science.aat7932}}.

\bibitem{liu2018simultaneously}
Te-Huan Liu, Bai Song, Laureen Meroueh, Zhiwei Ding, Qichen Song, Jiawei Zhou,
  Mingda Li, and Gang Chen.
\newblock Simultaneously high electron and hole mobilities in cubic boron-v
  compounds: Bp, bas, and bsb.
\newblock {\em Physical Review B}, 98\penalty0 (8):\penalty0 081203, 2018.
\newblock
  {\url{https://journals.aps.org/prb/abstract/10.1103/PhysRevB.98.081203}}.

\bibitem{bushick2020boron}
Kyle Bushick, Sieun Chae, Zihao Deng, John~T Heron, and Emmanouil Kioupakis.
\newblock Boron arsenide heterostructures: lattice-matched heterointerfaces and
  strain effects on band alignments and mobility.
\newblock {\em npj Computational Materials}, 6\penalty0 (1):\penalty0 3, 2020.
\newblock {\url{https://www.nature.com/articles/s41524-019-0270-4}}.

\bibitem{kim2016thermal}
Jaehyun Kim, Daniel~A Evans, Daniel~P Sellan, Owen~M Williams, Eric Ou, Alan~H
  Cowley, and Li~Shi.
\newblock Thermal and thermoelectric transport measurements of an individual
  boron arsenide microstructure.
\newblock {\em Applied Physics Letters}, 108\penalty0 (20):\penalty0 201905,
  2016.
\newblock {\url{https://aip.scitation.org/doi/10.1063/1.4950970}}.

\bibitem{shin2022high}
Jungwoo Shin, Geethal~Amila Gamage, Zhiwei Ding, Ke~Chen, Fei Tian, Xin Qian,
  Jiawei Zhou, Hwijong Lee, Jianshi Zhou, Li~Shi, Thanh Nguyen, Fei Han, Mingda
  Li, David Broido, Aaron Schmidt, Zhifeng Ren, and Gang Chen.
\newblock High ambipolar mobility in cubic boron arsenide.
\newblock {\em Science}, 377\penalty0 (6604):\penalty0 437--440, 2022.
\newblock {\url{https://www.science.org/doi/full/10.1126/science.abn4290}}.

\bibitem{yue2022high}
Shuai Yue, Fei Tian, Xinyu Sui, Mohammadjavad Mohebinia, Xianxin Wu, Tian Tong,
  Zhiming Wang, Bo~Wu, Qing Zhang, Zhifeng Ren, Jiming Bao, and Xinfeng Liu.
\newblock High ambipolar mobility in cubic boron arsenide revealed by transient
  reflectivity microscopy.
\newblock {\em Science}, 377\penalty0 (6604):\penalty0 433--436, 2022.
\newblock {\url{https://www.science.org/doi/full/10.1126/science.abn4727}}.

\bibitem{bernardi2014ab}
Marco Bernardi, Derek Vigil-Fowler, Johannes Lischner, Jeffrey~B Neaton, and
  Steven~G Louie.
\newblock Ab initio study of hot carriers in the first picosecond after
  sunlight absorption in silicon.
\newblock {\em Physical review letters}, 112\penalty0 (25):\penalty0 257402,
  2014.
\newblock
  {\url{https://journals.aps.org/prl/abstract/10.1103/PhysRevLett.112.257402}}.

\bibitem{giustino2017electron}
Feliciano Giustino.
\newblock Electron-phonon interactions from first principles.
\newblock {\em Reviews of Modern Physics}, 89\penalty0 (1):\penalty0 015003,
  2017.
\newblock
  {\url{https://journals.aps.org/rmp/abstract/10.1103/RevModPhys.89.015003}}.

\bibitem{sher1967resonant}
A~Sher and KK~Thornber.
\newblock Resonant electron-phonon scattering in polar semiconductors.
\newblock {\em Applied Physics Letters}, 11\penalty0 (1):\penalty0 3--5, 1967.
\newblock {\url{https://aip.scitation.org/doi/abs/10.1063/1.1754948}}.

\bibitem{ngai1972two}
KL~Ngai and EJ~Johnson.
\newblock Two-phonon deformation potential in insb.
\newblock {\em Physical Review Letters}, 29\penalty0 (24):\penalty0 1607, 1972.
\newblock
  {\url{https://journals.aps.org/prl/abstract/10.1103/PhysRevLett.29.1607}}.

\bibitem{alldredge1967role}
Gerald~P Alldredge and FJ~Blatt.
\newblock On the role of two-phonon processes in the energy relaxation of a
  heated-electron distribution.
\newblock {\em Annals of Physics}, 45\penalty0 (2):\penalty0 191--231, 1967.
\newblock
  {\url{https://www.sciencedirect.com/science/article/pii/0003491667901236}}.

\bibitem{lee2020ab}
Nien-En Lee, Jin-Jian Zhou, Hsiao-Yi Chen, and Marco Bernardi.
\newblock Ab initio electron-two-phonon scattering in gaas from next-to-leading
  order perturbation theory.
\newblock {\em Nature communications}, 11\penalty0 (1):\penalty0 1--7, 2020.
\newblock {\url{https://www.nature.com/articles/s41467-020-15339-0}}.

\bibitem{cheng2022high}
Peishi~S Cheng, Jiace Sun, Shi-Ning Sun, Alexander~Y Choi, and Austin~J
  Minnich.
\newblock High-field transport and hot-electron noise in gaas from
  first-principles calculations: Role of two-phonon scattering.
\newblock {\em Physical Review B}, 106\penalty0 (24):\penalty0 245201, 2022.
\newblock
  {\url{https://journals.aps.org/prb/abstract/10.1103/PhysRevB.106.245201}}.

\bibitem{hatanpaa2023two}
Benjamin Hatanp{\"a}{\"a}, Alexander~Y Choi, Peishi~S Cheng, and Austin~J
  Minnich.
\newblock Two-phonon scattering in nonpolar semiconductors: A first-principles
  study of warm electron transport in si.
\newblock {\em Physical Review B}, 107\penalty0 (4):\penalty0 L041110, 2023.
\newblock
  {\url{https://journals.aps.org/prb/abstract/10.1103/PhysRevB.107.L041110}}.

\bibitem{choi2021electronic}
Alexander~Y Choi, Peishi~S Cheng, Benjamin Hatanp{\"a}{\"a}, and Austin~J
  Minnich.
\newblock Electronic noise of warm electrons in semiconductors from first
  principles.
\newblock {\em Physical Review Materials}, 5\penalty0 (4):\penalty0 044603,
  2021.
\newblock
  {\url{https://journals.aps.org/prmaterials/abstract/10.1103/PhysRevMaterials.5.044603}}.

\bibitem{ponce2021first}
Samuel Ponc{\'e}, Francesco Macheda, Elena~Roxana Margine, Nicola Marzari,
  Nicola Bonini, and Feliciano Giustino.
\newblock First-principles predictions of hall and drift mobilities in
  semiconductors.
\newblock {\em Physical Review Research}, 3\penalty0 (4):\penalty0 043022,
  2021.
\newblock
  {\url{https://journals.aps.org/prresearch/abstract/10.1103/PhysRevResearch.3.043022}}.

\bibitem{bernardi2016first}
Marco Bernardi.
\newblock First-principles dynamics of electrons and phonons.
\newblock {\em The European Physical Journal B}, 89:\penalty0 1--15, 2016.
\newblock {\url{https://link.springer.com/article/10.1140/epjb/e2016-70399-4}}.

\bibitem{giannozzi2009quantum}
Paolo Giannozzi, Stefano Baroni, Nicola Bonini, Matteo Calandra, Roberto Car,
  Carlo Cavazzoni, Davide Ceresoli, Guido~L Chiarotti, Matteo Cococcioni,
  Ismaila Dabo, et~al.
\newblock Quantum espresso: a modular and open-source software project for
  quantum simulations of materials.
\newblock {\em Journal of physics: Condensed matter}, 21\penalty0
  (39):\penalty0 395502, 2009.

\bibitem{zhou2021perturbo}
Jin-Jian Zhou, Jinsoo Park, I-Te Lu, Ivan Maliyov, Xiao Tong, and Marco
  Bernardi.
\newblock Perturbo: A software package for ab initio electron--phonon
  interactions, charge transport and ultrafast dynamics.
\newblock {\em Computer Physics Communications}, 264:\penalty0 107970, 2021.
\newblock
  {\url{https://www.sciencedirect.com/science/article/pii/S0010465521000837}}.

\bibitem{ponce2018towards}
Samuel Ponc{\'e}, Elena~R Margine, and Feliciano Giustino.
\newblock Towards predictive many-body calculations of phonon-limited carrier
  mobilities in semiconductors.
\newblock {\em Physical Review B}, 97\penalty0 (12):\penalty0 121201, 2018.
\newblock
  {\url{https://journals.aps.org/prb/abstract/10.1103/PhysRevB.97.121201}}.

\bibitem{zhou2016ab}
Jin-Jian Zhou and Marco Bernardi.
\newblock Ab initio electron mobility and polar phonon scattering in gaas.
\newblock {\em Physical Review B}, 94\penalty0 (20):\penalty0 201201, 2016.
\newblock
  {\url{https://journals.aps.org/prb/abstract/10.1103/PhysRevB.94.201201}}.

\bibitem{kang2019basic}
Joon~Sang Kang, Man Li, Huan Wu, Huuduy Nguyen, and Yongjie Hu.
\newblock Basic physical properties of cubic boron arsenide.
\newblock {\em Applied Physics Letters}, 115\penalty0 (12):\penalty0 122103,
  2019.
\newblock {\url{https://aip.scitation.org/doi/full/10.1063/1.5116025}}.

\bibitem{buckeridge2019electronic}
John Buckeridge and David~O Scanlon.
\newblock Electronic band structure and optical properties of boron arsenide.
\newblock {\em Physical Review Materials}, 3\penalty0 (5):\penalty0 051601,
  2019.
\newblock
  {\url{https://journals.aps.org/prmaterials/pdf/10.1103/PhysRevMaterials.3.051601}}.

\bibitem{lyons2018impurity}
John~L Lyons, Joel~B Varley, Evan~R Glaser, Jaime~A Freitas~Jr, James~C
  Culbertson, Fei Tian, Geethal~Amila Gamage, Haoran Sun, Hamidreza Ziyaee, and
  Zhifeng Ren.
\newblock Impurity-derived p-type conductivity in cubic boron arsenide.
\newblock {\em Applied Physics Letters}, 113\penalty0 (25):\penalty0 251902,
  2018.
\newblock {\url{https://aip.scitation.org/doi/10.1063/1.5058134}}.

\bibitem{bushick2019band}
Kyle Bushick, Kelsey Mengle, Nocona Sanders, and Emmanouil Kioupakis.
\newblock Band structure and carrier effective masses of boron arsenide:
  Effects of quasiparticle and spin-orbit coupling corrections.
\newblock {\em Applied Physics Letters}, 114\penalty0 (2):\penalty0 022101,
  2019.
\newblock {\url{https://aip.scitation.org/doi/10.1063/1.5062845}}.

\bibitem{najafi2017super}
Ebrahim Najafi, Vsevolod Ivanov, Ahmed Zewail, and Marco Bernardi.
\newblock Super-diffusion of excited carriers in semiconductors.
\newblock {\em Nature communications}, 8\penalty0 (1):\penalty0 1--7, 2017.
\newblock {\url{https://www.nature.com/articles/ncomms15177}}.

\bibitem{ruzicka2010hot}
Brian~A Ruzicka, Shuai Wang, Lalani~K Werake, Ben Weintrub, Kian~Ping Loh, and
  Hui Zhao.
\newblock Hot carrier diffusion in graphene.
\newblock {\em Physical Review B}, 82\penalty0 (19):\penalty0 195414, 2010.
\newblock
  {\url{https://journals.aps.org/prb/abstract/10.1103/PhysRevB.82.195414}}.

\bibitem{choudhry2022persistent}
Usama Choudhry, Fengjiao Pan, Xing He, Basamat Shaheen, Taeyong Kim, Ryan
  Gnabasik, Geethal~Amila Gamage, Haoran Sun, Alex Ackerman, Ding-Shyue Yang,
  Zhifeng Ren, and Bolin Liao.
\newblock Persistent hot carrier diffusion in boron arsenide single crystals
  imaged by ultrafast electron microscopy.
\newblock {\em Matter}, 2022.
\newblock
  {\url{https://www.sciencedirect.com/science/article/pii/S2590238522005811}}.

\bibitem{holstein1959theory}
T~Holstein.
\newblock Theory of ultrasonic absorption in metals: the collision-drag effect.
\newblock {\em Physical review}, 113\penalty0 (2):\penalty0 479, 1959.
\newblock {\url{https://journals.aps.org/pr/abstract/10.1103/PhysRev.113.479}}.

\bibitem{kocevar1980multiphonon}
P~Kocevar.
\newblock Multiphonon scattering.
\newblock {\em Physics of Nonlinear Transport in Semiconductors}, pages
  167--174, 1980.
\newblock
  {\url{https://link.springer.com/chapter/10.1007/978-1-4684-3638-9_7}}.

\bibitem{ngai1974carrier}
KL~Ngai.
\newblock Carrier-two phonon interaction in semiconductors.
\newblock In {\em Proceedings of the Twelfth International Conference on the
  Physics of Semiconductors: July 15--19, 1974 Stuttgart}, pages 489--498.
  Springer, 1974.
\newblock
  {\url{https://link.springer.com/chapter/10.1007/978-3-322-94774-1_83}}.

\bibitem{fasol1990continuous}
G~Fasol, W~Hackenberg, HP~Hughes, K~Ploog, E~Bauser, and H~Kano.
\newblock Continuous-wave spectroscopy of femtosecond carrier scattering in
  gaas.
\newblock {\em Physical Review B}, 41\penalty0 (3):\penalty0 1461, 1990.
\newblock
  {\url{https://journals.aps.org/prb/abstract/10.1103/PhysRevB.41.1461}}.

\end{thebibliography}

\end{document}